\documentclass[10pt,final,journal,letter,oneside,twocolumn]{IEEEtran}

\usepackage[noadjust]{cite}
\usepackage{srcltx}
\usepackage{psfrag}
\usepackage{epsfig}
\usepackage[tbtags,sumlimits,nointlimits,reqno]{amsmath}
\usepackage{amssymb}
\usepackage{xcolor}
\usepackage{float}
\usepackage{subfigure} 
\usepackage{soul}
\usepackage{footnote}

\usepackage{graphicx}

\setstcolor{red}

\usepackage{showkeys}   

\begin{document}

\title{Unveiling Magnetic Dipole Radiation in Phase-Reversal Leaky-Wave Antennas}

\author{%
       Shulabh~Gupta,~\IEEEmembership{Member,~IEEE,} Li Jun Jiang,~\IEEEmembership{Senior Member,~IEEE,} and~Christophe~Caloz,~\IEEEmembership{Fellow,~IEEE}
\thanks{S. Gupta and L. J. Jiang are with the Electrical and Electronic Engineering Department of University of Hong Kong, China (SAR).}
\thanks{C. Caloz is with the Department of Electrical
Engineering, PolyGrames Research Center, \'{E}cole Polytechnique de Montr\'{e}al,
Montr\'{e}al, QC, Canada.}
}
\markboth{IEEE Antenna and Wireless Propagation Letters 2013}{Shell \MakeLowercase{\textit{et al.}}: Bare Demo of IEEEtran.cls for
Journals}

\maketitle

\begin{abstract}
The radiation principle of travelling-wave type phase-reversal antennas is explained in details, unveiling the presence of magnetic-dipole radiation in addition to well-known electric dipole radiation. It is point out that such magnetic-dipole radiation is specific to the case of traveling-wave phase-reversal antennas whereas only electric-dipole radiation exists in resonant-type phase-reversal antennas. It is shown that a phase-reversal travelling-wave antenna alternately operates as an array of magnetic dipoles and an array of electric-dipoles during a time-harmonic period. This radiation mechanism is confirmed through both full-wave and experimental results.\end{abstract}


%
\IEEEpeerreviewmaketitle
%
\section{Introduction}

Periodic leaky-wave structures are a versatile class of travelling-wave antennas capable of frequency-controlled beam steering, offering high directivity and simple feeding mechanism \cite{Leaky_Book}. With the advent of composite right/left-handed (CRLH) leaky-wave antennas capable of full-space scanning, from backward to forward including broadside directions, there has been a renewed interest in such structures \cite{Leaky_jackson}. Full-scpae scanning was made possible by suppressing the open stopband and equalizing the impedance around broadside \cite{Leaky_jackson}\cite{Otto_Leaky}\cite{Caloz-MTM-Book}. Following these advances, periodic leaky-wave antennas have become most attractive as low-cost alternative to phased array antennas \cite{Balanis-Antenna-Book}.

One approach to achieve a full-space beam scanning in leaky-wave structures, is to use the phase-reversal technique. This technique involves introducing a $180^\circ$ phase shift along the axis of the structure, so that $\lambda_g/2$-spaced periodically elements get excited in phase and therefore radiate. This technique is common in slot-waveguide antenna arrays where the required extra $180^\circ$ phase-shift is induced by proper location of the slots along the waveguide aperture \cite{Silver_Book}. A transmission-line counterpart of this structure is the Sterba-curtain antenna, which utilizes periodic transmission-line cross-overs, placed $\lambda_g/2$ apart, to produce electric-dipole radiation from the cross-overs \cite{Sterba_Patent}. However, these waveguide and transmission-line antennas were \textit{resonant} in nature and therefore featured a small impedance-matching bandwidth product \cite{Yang_PR_WideBW},\cite{Krauss_Book}. This phase reversal concept was later extended to a \textit{travelling-wave} configurations, thereby both improving the impedance-matching bandwidth performance and providing full-space scanning. This configuration uses periodic phase-reversal radiating elements interconnected by balanced transmission line sections \cite{Yang_FullSpacePR}.

\begin{figure}[htbp]
\begin{center}
\subfigure[]{
\psfrag{a}[c][c][1]{$=$}
\psfrag{+}[c][c][0.8]{$+$}
\psfrag{-}[c][c][0.8]{$-$}
\psfrag{d}[c][c][0.8]{$c_1$}
\psfrag{e}[c][c][0.8]{$c_2$}
\psfrag{g}[c][c][0.8]{$\lambda/2$}
\psfrag{x}[c][c][0.8]{$\mathbf{m}$}
\psfrag{y}[c][c][0.8]{$-\mathbf{m}$}
\psfrag{b}[c][c][0.8]{\shortstack{mutually cancelling \\magnetic dipoles}}
\psfrag{c}[c][c][0.8]{conventional differential lines}
\includegraphics[width=0.98\columnwidth]{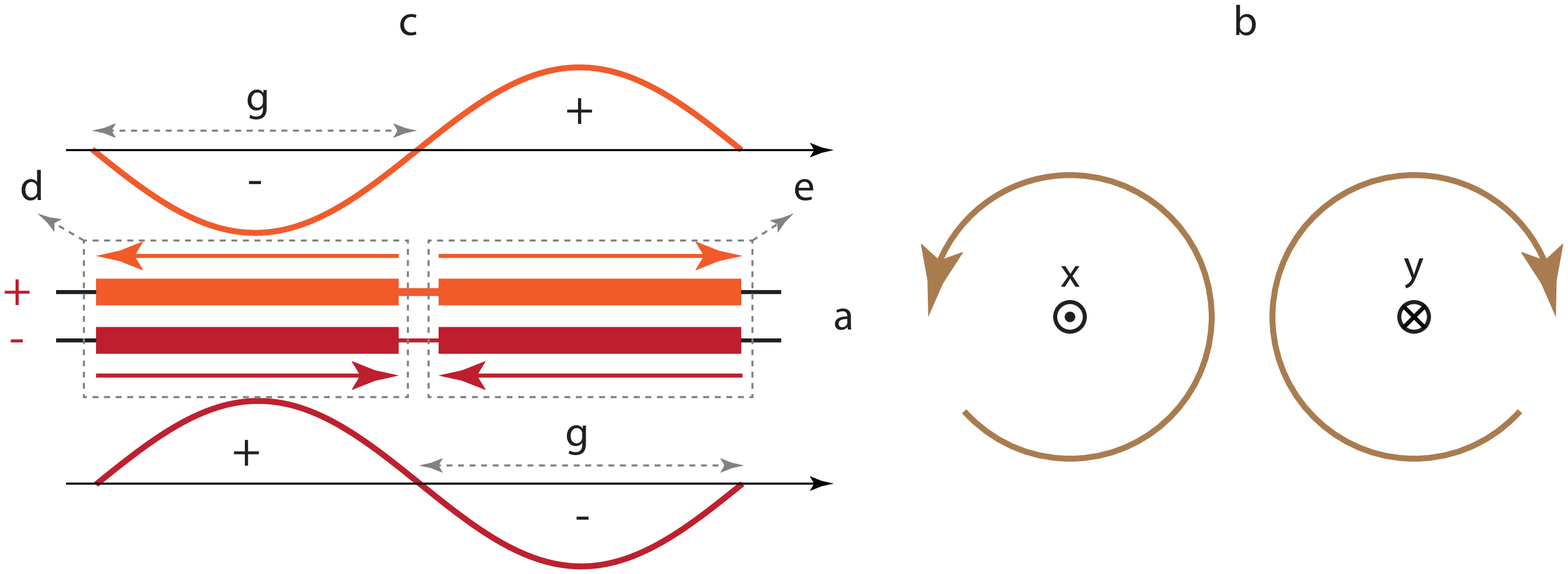}}
\subfigure[]{
\psfrag{a}[c][c][1]{$=$}
\psfrag{+}[c][c][0.8]{$+$}
\psfrag{-}[c][c][0.8]{$-$}
\psfrag{x}[c][c][0.8]{$\mathbf{m}$}
\psfrag{b}[c][c][0.8]{\shortstack{radiating magnetic dipoles}}
\psfrag{c}[c][c][0.8]{phase-reversal  antenna at $t=0$}
\includegraphics[width=0.98\columnwidth]{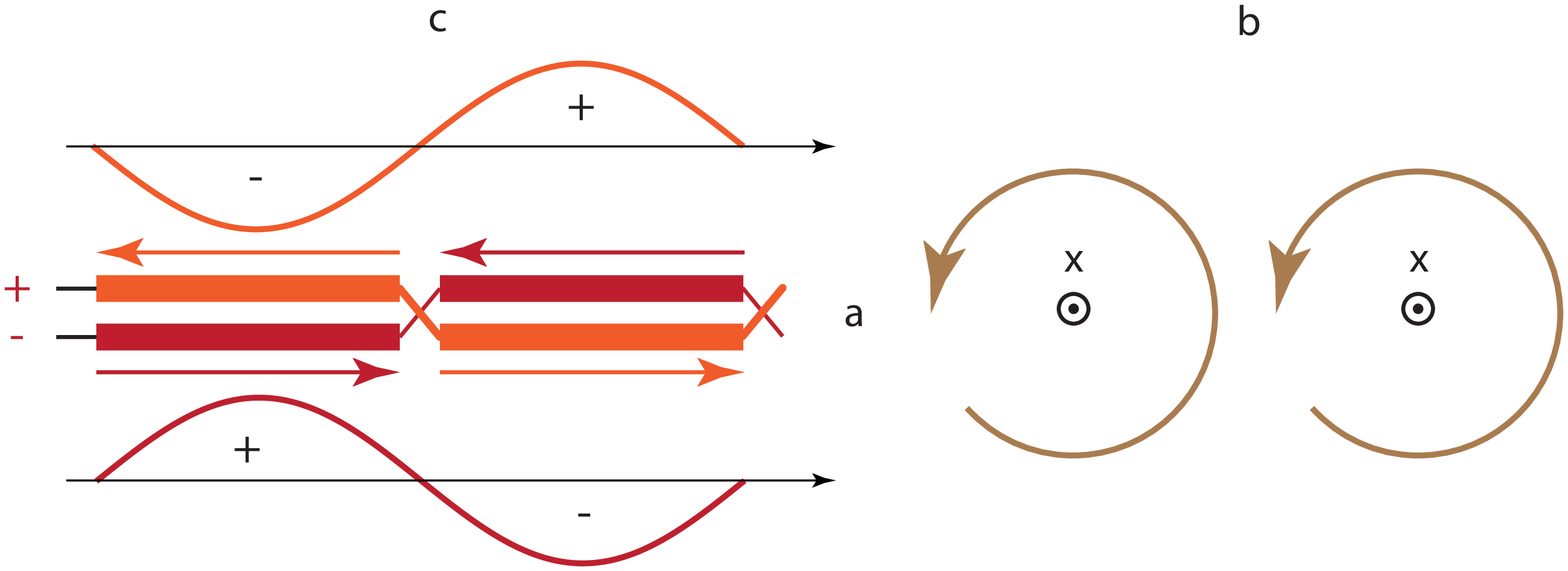}}
\subfigure[]{
\psfrag{a}[c][c][1]{$=$}
\psfrag{+}[c][c][0.8]{$+$}
\psfrag{-}[c][c][0.8]{$-$}
\psfrag{x}[c][c][0.8]{$\mathbf{p}$}
\psfrag{b}[c][c][0.8]{\shortstack{radiating electric dipole}}
\psfrag{c}[c][c][0.8]{phase-reversal antenna at $t=T/4$}
\includegraphics[width=0.98\columnwidth]{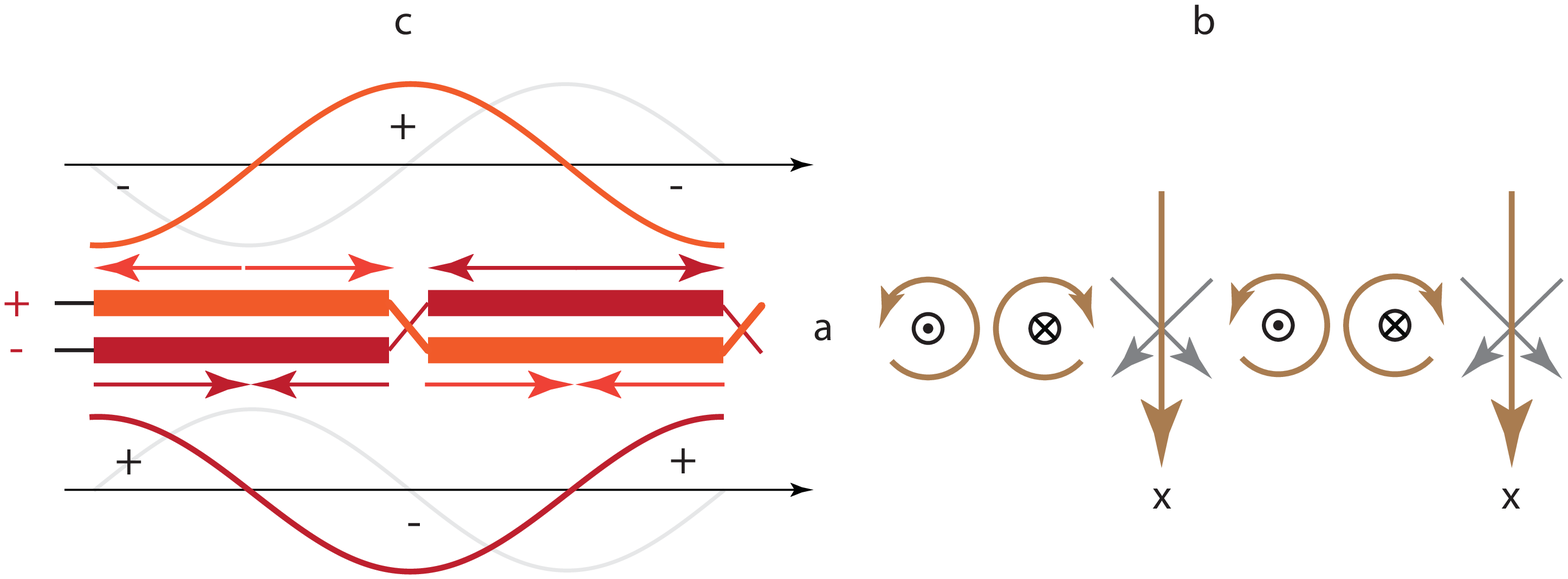}}
\caption{Principle of phase-reversal antenna explained in terms of current distributions in two adjacent unit cells. a) Differential transmission line without phase reversal. b) Phase reversed differential lines at $t=0$ and, c) at $t = T/4$, where $T = 1/f_0$ is the time period of an assumed time-harmonic signal of frequency $f_0$.}\label{Fig:PRprinciple} 
\end{center}
\end{figure}

\begin{figure*}[htbp]
\begin{center}
\subfigure[]{\psfrag{z}[c][c][0.7]{$z$}
\psfrag{y}[c][c][0.8]{$y$}
\psfrag{x}[c][c][0.8]{$x$}
\psfrag{e}[c][c][0.8]{$p=\lambda_g/2$~at~$f_0$}
\psfrag{f}[c][c][0.8]{$\Delta\ell$}
\psfrag{a}[c][c][0.8]{$w$}
\psfrag{b}[c][c][0.8]{$g$}
\psfrag{c}[c][c][0.8]{$s$}
\psfrag{d}[c][c][0.8]{$h$}
\includegraphics[width=2\columnwidth]{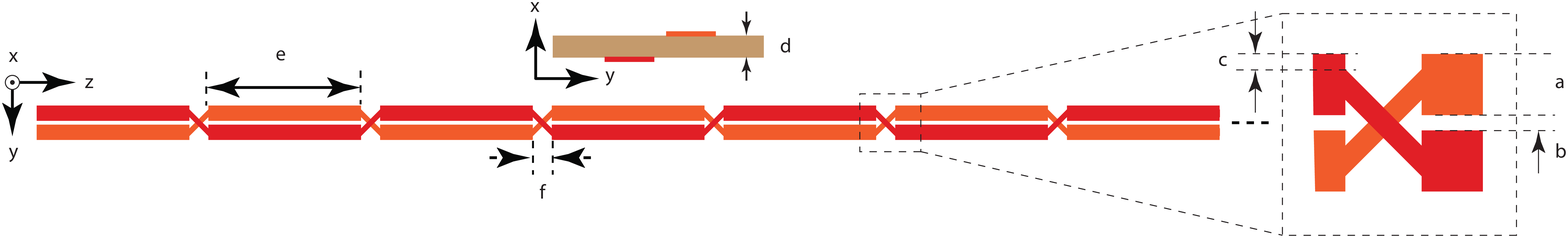}}
\subfigure[]{
\psfrag{a}[c][c][0.8]{\shortstack{M-dipole $G_\theta(\phi)$ \\ $\theta=90^\circ$ ($x-y$ plane)}}
\psfrag{b}[c][c][0.8]{\shortstack{E-dipole $G_\phi(\phi)$ \\ $\theta=90^\circ$($x-y$ plane)}}
\psfrag{c}[c][c][0.8]{\shortstack{M-dipole $G_\theta(\theta)$ \\ $\phi=90^\circ$ ($y-z$ plane)}}
\psfrag{d}[c][c][0.8]{\shortstack{E-dipole $G_\phi(\theta)$ \\ $\phi=0^\circ$ ($x-z$ plane)}}
\psfrag{p}[c][c][0.8]{$G_z(f < f_0)$}
\psfrag{q}[c][c][0.8]{\shortstack{$G_z(f=f_0)$}}
\psfrag{r}[c][c][0.8]{$G_z(f > f_0)$}
\psfrag{u}[c][c][0.8]{$G_y(f < f_0)$}
\psfrag{v}[c][c][0.8]{\shortstack{ $G_y(f=f_0)$}}
\psfrag{w}[c][c][0.8]{$G_y(f > f_0)$}
\psfrag{l}[l][c][0.8]{$f < f_0$}
\psfrag{m}[l][c][0.8]{$f = f_0$}
\psfrag{n}[l][c][0.8]{$f > f_0$}
\psfrag{z}[c][c][1]{$x$}
\psfrag{y}[c][c][1]{$z$}
\psfrag{x}[c][c][1]{$x$}
\includegraphics[width=2\columnwidth]{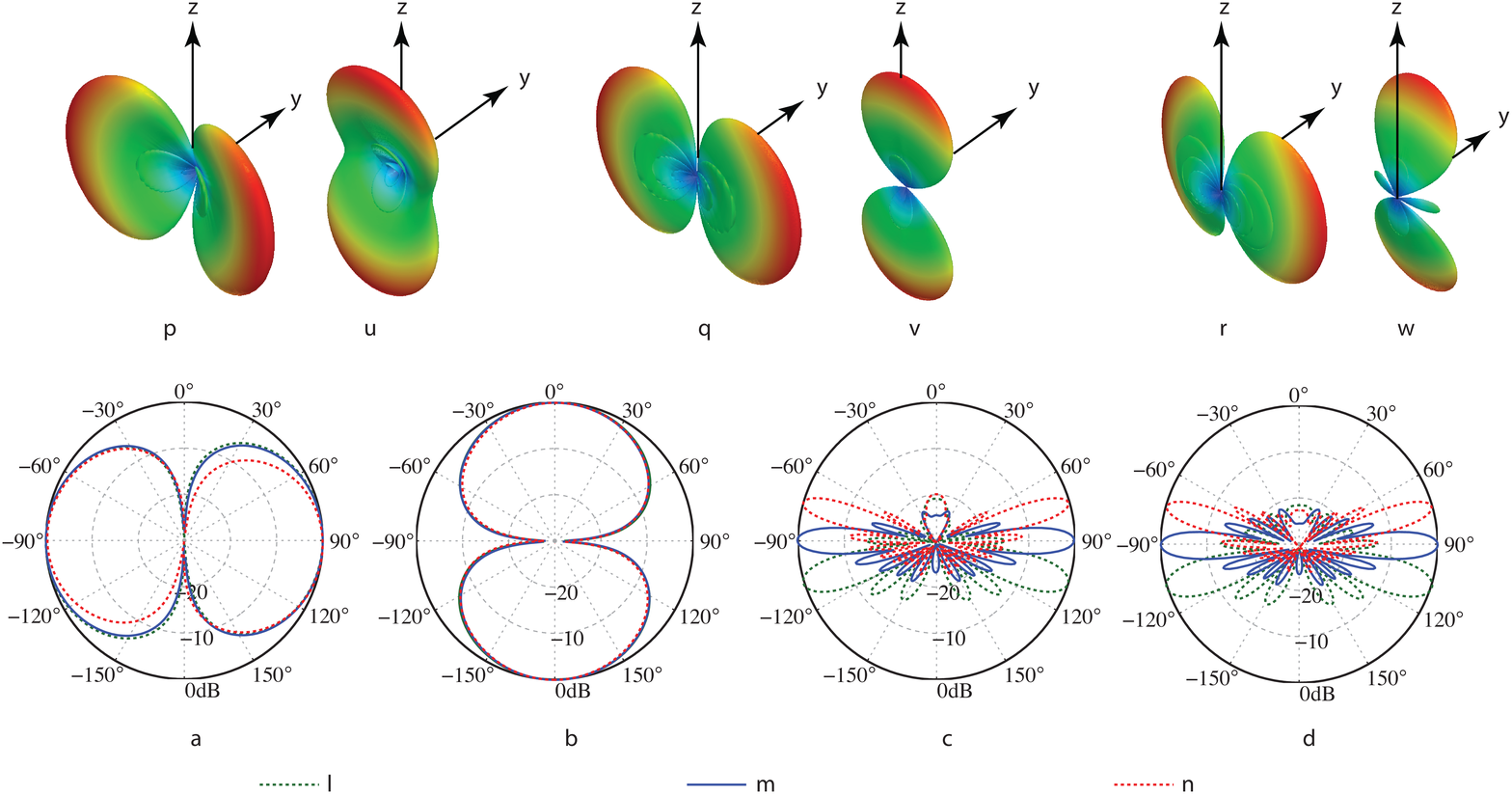}}
\caption{Planar transmission line implementation of a phase-reversal leaky-wave antenna and its typical radiation pattern characteristics \cite{Yang_FullSpacePR}. a) Antenna layout. b) 3D polar plot and the corresponding 2D radiation patterns in three principal cuts, computed using FEM-HFSS.} \label{Fig:HFSS}
\end{center}
\end{figure*}

\begin{figure*}[htbp]
\begin{center}
\subfigure[]{
\psfrag{z}[c][c][1]{$z$}
\psfrag{y}[c][c][1]{$y$}
\psfrag{x}[c][c][1]{$x$}
\psfrag{a}[l][c][0.8]{50$\Omega$ to $100\Omega$ microstrip to differential  transition}
\psfrag{b}[r][c][0.8]{$100\Omega$ termination}
\psfrag{c}[c][c][0.8]{$\ell = 34$~cm (20 cross-overs)}
\includegraphics[width=2\columnwidth]{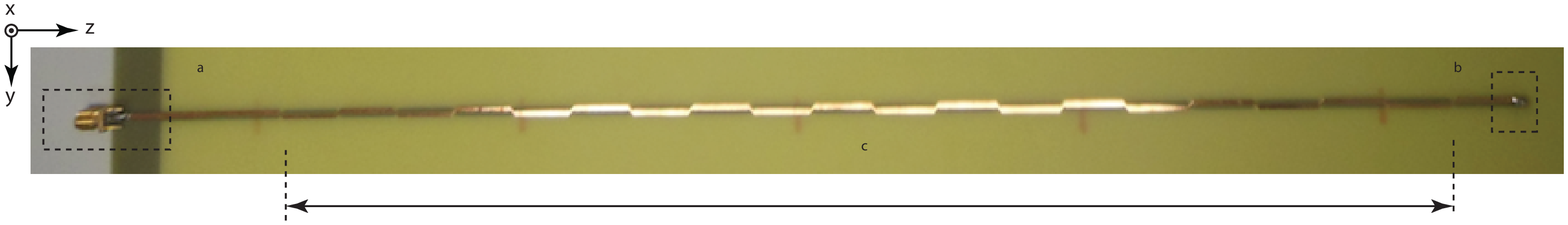}}
\subfigure[]{
\psfrag{d}[c][c][0.8]{$S_{11}$ (dB)}
\psfrag{x}[l][c][0.7]{Measured}
\psfrag{y}[l][c][0.7]{FEM-HFSS}
\psfrag{a}[c][c][0.8]{frequency $f$ (GHz)}
\includegraphics[width=0.6\columnwidth]{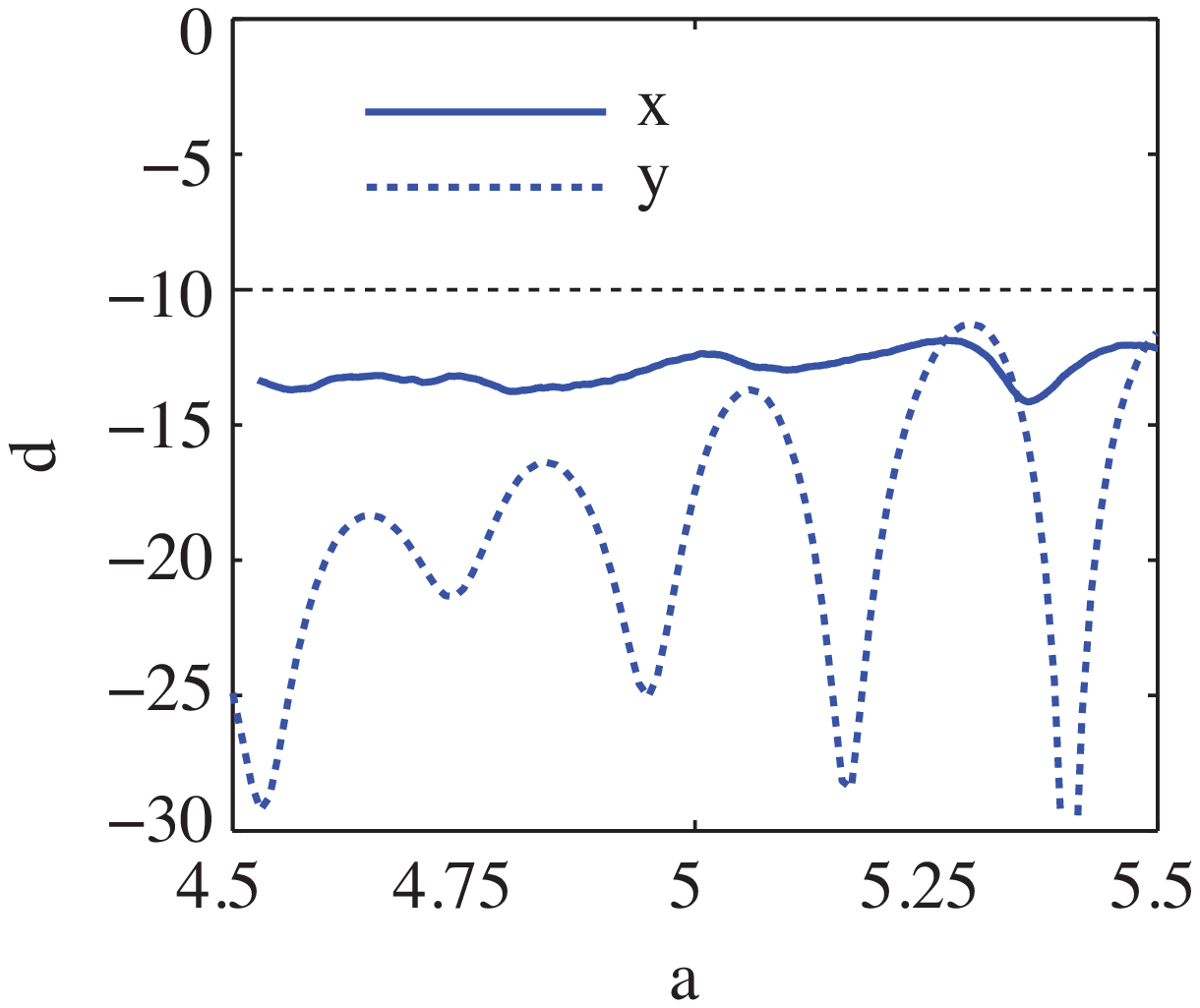}}
\subfigure[]{
\psfrag{e}[c][c][0.8]{measured $G_\phi(\theta, \phi)$}
\psfrag{f}[c][c][0.8]{measured $G_\theta(\theta, \phi)$}
\psfrag{b}[c][c][0.8]{Azimuth angle $\phi \times \pi$ rads}
\psfrag{c}[c][c][0.8]{Elevation angle $\theta \times \pi$ rads}
\includegraphics[width=1.4\columnwidth]{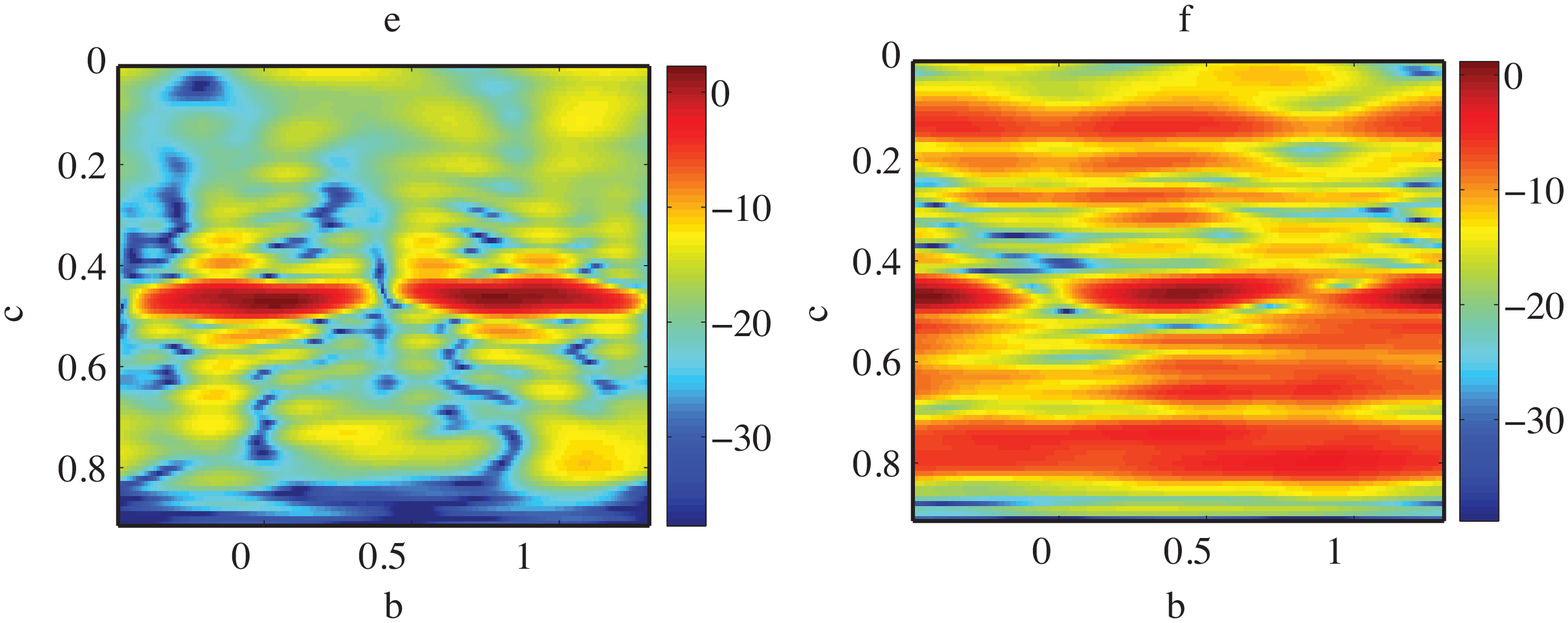}}
\subfigure[]{
\psfrag{a}[c][c][0.8]{\shortstack{E-dipole scanning \\$\phi= 0^\circ$ ($x-z$ plane)}}
\psfrag{b}[c][c][0.8]{\shortstack{M-dipole scanning\\$\phi= 90^\circ$ ($y-z$ plane)}}
\psfrag{c}[c][c][0.8]{\shortstack{E- and M-dipole\\$\theta= 90^\circ$ ($x-y$ plane)}}
\psfrag{e}[l][c][0.8]{$G_\theta$ (dB)}
\psfrag{d}[l][c][0.8]{$G_\phi$ (dB)}
\psfrag{x}[l][c][0.7]{Measured}
\psfrag{y}[l][c][0.7]{FEM-HFSS}
\includegraphics[width=2\columnwidth]{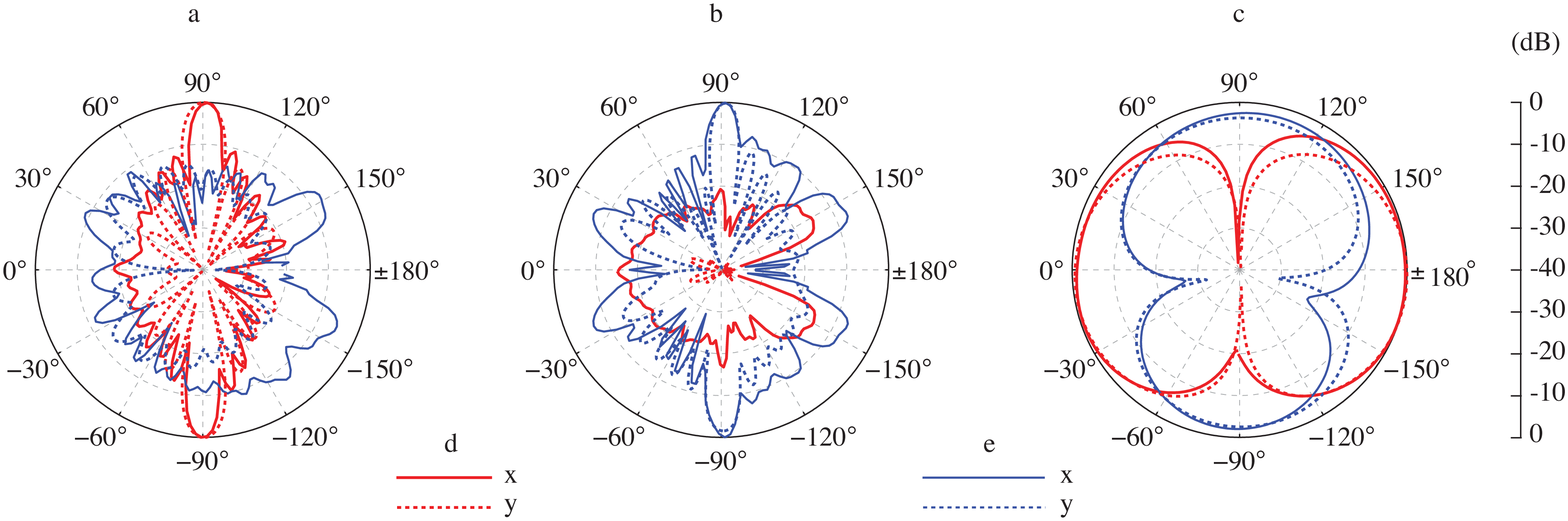}}
\caption{Fabricated phase reversal antenna and measured results. a) Photograph of the antenna. b) Measured S-parameters. c) Measured antenna gain and d) 2D radiation patterns in all three principal planes, for the transition frequency $f_0 = 4.8$~GHz. The layout parameters are: $p=16.51$~mm, $g=0.8$~mm, $\Delta\ell = 1.27$~mm, $w = 1.27$~mm, $s = 0.254$~mm and number of cross-overs $N = 20$.} \label{Fig:MeasuredPR}
\end{center}
\end{figure*}

While the radiation mechanism of traditional transmission-line type phase-reversal antennas is well understood in resonant configurations and extensively discussed in the literature \cite{Hessel_Book}, \textit{travelling-wave} phase-reversal antennas have not been exhaustively explored. Specifically, they have been described in terms of electric-dipole radiation only \cite{Yang_FullSpacePR}. In this work, the radiation mechanism of a travelling-wave type phase reversal antenna is revisited in details and unveils the presence of \textit{magnetic-dipole radiation} operating in conjunction with electric-dipole radiation. This dual radiation mechanism is specific to travelling-wave type phase-reversal antennas and does not exist in their resonant counterpart. They fall in the category of magneto-electric dipole antennas, where the electric- and magnetic-dipoles radiate together from a single radiating structure \cite{Luk_EM}, \cite{Rennings_Leaky}. The detailed principle of the phase-reversal antennas is explained next.


\section{Principle of a Phase-reversal Leaky-wave Antenna}

Consider a matched transmission line consisting of two closely-spaced conductors of length $\ell = \lambda$ at a frequency $f_0$, as shown in Fig.~\ref{Fig:PRprinciple}(a). This transmission line can be considered as two half-wavelength long lines connected in tandem via an ideal transmission line of zero length. Such a line supports a differential mode, since the current distributions on the top and bottom conductors are mirror images of each other. In such a configuration, the pair of parallel half-wavelength transmission lines may be seen as one-wavelength long current loops, $c_1$ and $c_2$, forming two adjacent magnetic dipoles, $\mathbf{m}$, as illustrated in Fig.~\ref{Fig:PRprinciple}(a). These two magnetic dipoles are \textit{anti-parallel}, and therefore do not radiate since they produce mutually cancelling far fields.

Fig.~\ref{Fig:PRprinciple}(b) shows a modified configuration the differential line, where a $\pi$-phase shift, commonly called phase-reversal, is introduced by crossing the connections between the conductor pairs. Consider a time-harmonic signal, of period $T$. At a certain time instant $t=0$, the two adjacent magnetic dipoles become \textit{parallel} to each other, as a result of phase reversal which produces magnetic-dipole far-field radiation. At that instant, the currents in the cross-over region are minimal and out-of-phase, and therefore do not radiate. The phase-reversal antenna may thus be seen as a series-fed array of magnetic-dipole antennas at time $t=0$.

At a later time instant $t=T/4$, the current distribution on the phase-reversal antenna changes as shown in Fig.~\ref{Fig:PRprinciple}(c). In this case, the current currents are maximal and in-phase at the cross-over region, leading to the formation of a radiating electric dipole moment, $\mathbf{p}$, perpendicular to the axis of the lines. While part of the current distribution on the transmission lines may be mapped to small current loops, the corresponding adjacent magnetic dipole moments are out-of-phase and therefore do not radiate. The phase-reversal antenna may thus be seen as a series-fed array of electric-dipole antennas, at time time $t=T/4$ \cite{Yang_FullSpacePR}.


\section{Radiation Characteristics and Results}


Based on the above explanation, it is clear that a phase-reversal antenna is based on two distinct radiation mechanisms. It alternates between magnetic-dipole radiation and orthogonally-polarized electric radiation, with a combination of the two in between these two instants. To further investigate the radiation properties of a phase-reversal leaky-wave antennas, designed in planar transmission line technology, is shown in Fig.~\ref{Fig:HFSS}(a). The antenna consists of two conducting strips on each side of a dielectric substrate, with a strip width $w$ and offset $g$, to form a transmission line of characteristic impedance $Z_0$. The cross-over region is realized using a small transmission line section of length $\Delta \ell$ with an impedance $Z_c$ such that $Z_c \ne Z_0$. This cross-over region with an impedance step acts as a matching element to suppress the well-known open stopband, typical of periodic structures \cite{Leaky_Book}, as proposed and demonstrated in \cite{Yang_FullSpacePR}. Under the condition of complete stop band suppression, there is a zero phase shift across the unit cell at the design frequency, $f_0$, which is sometimes also referred to as the transition frequency of the antenna, corresponding to peak radiation at broadside.

Typical radiation patterns for the phase-reversal leaky-wave antenna at the transition frequency are shown in Fig.~\ref{Fig:HFSS}(b), shown also in a 3D polar plot for better visualization. The electric and magnetic dipole radiation components are clearly apparent in the $x-y$ plane. While a $y-$directed electric dipole, from the cross-overs, produces a maximum along the $x-$axis, the $x-$directed and orthogonally polarized magnetic dipole produces a null in this direction. Similarly a magnetic dipole maximum along the $y$-axis aligns with the minimum of the $y-$directed electric dipole. In the other two planes, the magnetic-dipole component of the phase-reversal antenna with a maximum along the $y-$direction forms a directive beam in the $y-z$ plane, and its orthogonally-polarized electric-dipole component with a maximum along the $x-$direction forms a directive beam in the $x-z$ plane. These two planes are also the frequency scanning planes where the magnetic-dipole array and the electric-dipole array scans in the $y-z$ and the $x-z$ planes, respectively as seen in Fig.~\ref{Fig:HFSS}(b).

To further confirm the radiation characteristics of the phase-reversal leaky-wave antenna, a prototype is built on an FR4 substrate ($\varepsilon_r = 4.4$ and $h=0.8$~mm), as shown in Fig.~\ref{Fig:MeasuredPR}(a). It is fed through a multi-stage quarter-wavelength transformer to excite the differential mode of the phase-reversal antenna using a 50~$\Omega$ microstrip feed \cite{Yang_FullSpacePR}. Its measured return loss $S_{11}$ is shown in Fig.~\ref{Fig:MeasuredPR}(b) showing a satisfactory matching, with $S_{11}<-10$~dB within the band of interest. Fig.~\ref{Fig:MeasuredPR}(c) shows the measured radiation map of the antenna gain at $f_0=4.8$~GHz, for its two orthogonal components $G_\phi$ and $G_\theta$. The corresponding radiation patterns in three principal plane cuts are plotted in Fig.~\ref{Fig:MeasuredPR}(d). As expected, two distinct sets of radiation patterns, corresponding to the magnetic-dipole array and an electric-dipole arrays are clearly seen in all planes with peak gains of 1.2~dB and 2.3~dB, respectively. A reasonable agreement is also observed between the measured and the full-wave simulated results, with a stronger agreement at higher gain values, due to increased measurement sensitivity.

\section{Conclusions}

The radiation principle of a travelling-wave type phase reversal antenna has been explained in details, unveiling the presence of magnetic-dipole radiation which exist in addition to the electric-dipole radiation. This is in contrast to the resonant configuration where only electric-dipole radiation exists. It has been shown that the radiation characteristics of a phase-reversal antenna alternates between that of an array of magnetic dipoles and of an array of electric dipoles. While the crossover regions provide the electric-dipole radiation, the balanced transmission line between the cross-overs form a wavelength-long current loop at the designed frequency, representing the magnetic dipole radiation. This radiation principle has been confirmed using both full-wave and experimental results. This insight into the phase-reversal type antenna highlights the rich radiation characteristics of phase-reversal antennas compared to what was previously understood, and may lead to efficient and multifunctional leaky-wave antennas. 

\section*{Acknowledgement}
Authors would like to thank the Radio-frequency Radiation Research Laboratory at the Chinese University of Hong Kong for their help in radiation pattern measurements. This work was supported by HK ITP/026/11LP, HK GRF 711511, HK GRF 713011, HK GRF 712612, and NSFC 61271158.

\bibliographystyle{IEEEtran}
\bibliography{Gupta_PhaseReversal_EM_TAP_2013_Ref}
\end{document}